\documentstyle[12pt,epsf,epsfig,wrapfig,amsfonts,bm]{article}
\begin{document}

\begin{center}
{\bfseries SPIN EFFECTS FOR NEUTRINOS AND ELECTRONS MOVING IN DENSE MATTER}

\vskip 5mm A.V. Grigoriev $^{1}$, A.M. Savochkin $^{2}$, \underline{A.I.
Studenikin}$^{2 \dag}$ and A.I. Ternov$^{3}$

\vskip 5mm {\small (1) {\it Skobeltsin Institute of Nuclear Physics, Moscow State
University}\\
(2) {\it Department of Theoretical Physics, Moscow State University
}\\
(3) {\it Department of Theoretical Physics, Moscow Institute for Physics and Technology
}\\
$\dag$ {\it E-mail: studenik@srd.sinp.msu.ru}}
\end{center}

\vskip 5mm
\begin{abstract}
We shortly summarize the present status of neutrino magnetic moment studies (theory and experiment). Then we discuss
the quasiclassical treatment of neutrino spin evolution in matter. After that we come to the quantum approach to
description of neutrino and electron motion in matter on the basis of the quantum wave equations exact solutions method
with special focus on spin effects.
\end{abstract}

This paper is devoted to the problem of neutrino and electron motion in a dense matter with special focus on the spin
phenomena.

It has been proven in recent oscillation experiments that neutrino has nonzero
mass. Therefore, the Dirac neutrino should have nontrivial electromagnetic
properties, in particular, nonzero magnetic moment. It is also well known
\cite{FujShr80} that in the minimally extended Standard Model with
$SU(2)$-singlet right-handed neutrino the one-loop radiative correction
generates neutrino magnetic moment which is proportional to the neutrino mass
$\mu_{\nu}={3 \over {8 \sqrt{2}\pi^{2}}}eG_{F}m_{\nu}=3 \times 10^{-
19}\mu_{0}\bigg({m_{\nu} \over {1 \mathrm{eV}}}\bigg)$, where $\mu_{0}=e/2m$ is
the Bohr magneton, $m_{\nu}$ and $m$ are the neutrino and electron masses.
There are also models (see \cite{KBMVVFY76-87}) in which much large values for
magnetic moment of neutrino are predicted.

The LEP data require that the number of light neutrinos coupling to Z boson is
exactly three, whereas any additional neutrino, if this particle exist, must be
heavy. In light of this opportunity we considered the neutrino magnetic moment
for various ratios of particles masses. We have obtained
\cite{DvoStuPRD04DvoStuJETP04} values of the neutrino magnetic moment for light
(for this particular case see also \cite{FujShr80,CarBerVidZep00}),
intermediate and heavy massive neutrino: $ 1)\ \mu_{\nu}=
    {\frac{eG_{F}}{4\pi^{2}\sqrt{2}}}m_{\nu}
    {\frac{3(2-7a+6a^{2}-2a^{2}\ln a-a^{3})}{4(1-a)^{3}}}$,
    for $m_\nu\ll m_\ell\ll M_W,$
2) $\ \mu_{\nu}={\frac{3eG_{F}}{8\pi^{2}\sqrt{2}}}m_{\nu}
    \left\{
    1+{\frac{5}{18}}b
    \right\}$, \ for   $m_\ell\ll m_\nu\ll M_W,$
3) $\mu={\frac{eG_{F}}{8\pi^{2}\sqrt{2}}}m_{\nu}$, for $m_\ell\ll M_W\ll
m_\nu,$ where $\ \ a=\big( {m_{l} \over M_{W}}\big)^{2}$ and $b=\big( {m_{\nu}
\over M_{W}}\big)^{2}$. It should be also mentioned that the neutrino magnetic
moment can be affected by the external environment. In particular, the value of
the neutrino magnetic moment can be significantly shifted by the presence of
strong external magnetic fields \cite{BorZhuKurTer85} (see also
\cite{EgoLikStuEPP99,Nie03}).

So far, solar neutrino experiments set a limit on the neutrino magnetic moment on the level of $\mu_{\nu_{e}} \leq 1.5
\times 10^{-10}$\cite{BeaVogPRL99LiuPRL04}. More stringent constraint $\mu_{\nu_{e}} \leq 5.8 \times 10^{-11}$ has been
provided by the GEMMA accelerator experiment \cite{StaPAN07}. The constraint from astrophysical considerations (the red
giants cooling) is  $\mu_{\nu_{e}} \leq 3 \times 10^{-12}$ \cite{RafPR99}. There are also constraints coming from
estimations based on supernovae core collapse and primordial nucleosynthesis: $\mu_{\nu} \leq 3 \times 10^{-10}$ (see
\cite{D'Olivo99-00} and references therein).

Developing of the theory of neutrino spin properties in an external environment
we have evaluated the Loretz invariant approach to the neutrino spin evolution
that was based on the proposed generalized Bargmann-Michel-Telegdi equation
\cite{EgoLobStuPLB_00}. Within the developed Lorentz invariant approach it is
also possible  to find the solution for the neutrino spin evolution problem for
a general case when  the neutrino is subjected to general types of
non-derivative interactions with external fields \cite{DvoStuJHEP02}. These
interactions are given by the Lagrangian
\begin{equation}
-{\cal L}=g_{s}s(x){\bar \nu}\nu+ g_{p}{\pi}(x) {\bar \nu}\gamma^{5}\nu+
g_{v}V^{\mu}(x){\bar \nu}\gamma_{\mu}\nu+ g_{a}A^{\mu}(x){\bar
\nu}\gamma_{\mu}\gamma^{5}\nu+ {{g_{t}}\over{2}}T^{\mu\nu}{\bar
\nu}\sigma_{\mu\nu}\nu+ {{g^{\prime}_{t}}\over{2}} \Pi^{\mu\nu}{\bar
\nu}\sigma_{\mu\nu}\gamma_{5}\nu,
 \end{equation} where $s, \pi,
V^{\mu}=(V^{0}, {\bf V}), A^{\mu}=(A^{0}, {\bf A}), T_{\mu\nu}=({\bf a}, {\bf
b}), \Pi_{\mu\nu}=({\bm c}, {\bm d})$ are the scalar, pseudoscalar, vector,
axial-vector, tensor and pseudotensor fields, respectively. For the
corresponding spin evolution equation we have found
\begin{equation}\label{S_eq_gen}
\begin{array}{c}  {{d{\bf S} \over dt}}= 2g_{a}\left\{ A^{0}[{\bf S}
\times{\bm \beta}]- {{({\bf A}{\bm \beta})[{\bf S} \times{\bm
\beta}]}\over{1+{\gamma}^{-1}}} - {1 \over \gamma}[{\bf S}\times{\bf A}]
\right\}
 +2g_{t}\left\{ [{\bf S}\times{\bf b}]-
{{({\bm \beta}{\bf b})[{\bf S}\times{\bm \beta}]}\over{1+{\gamma}^{-1}}} +
[{\bf S}\times[{\bf a}\times{\bm \beta}]] \right\} \\
+ 2ig^{\prime}_{t}\left\{ [{\bf S}\times{\bf c}]- {{({\bm \beta}{\bf c})[{\bf
S}\times{\bm \beta}]}\over{1+{\gamma}^{-1}}}- [{\bf S}\times[{\bf d}\times{\bm
\beta}]] \right\}.
\end{array}
\end{equation}
This is a rather general equation for the neutrino spin evolution that can be
also used for description of neutrino spin oscillations  in different
environments such as moving and polarized matter with external electromagnetic
fields (see \cite{StuLatuile04, StuPAN07}).

Considering the neutrino spin evolution within the quasi-classical treatment on
the basis of the above mentioned  generalized Bargmann-Michel-Telegdi equation,
we have predicted \cite{LobStudPLB03} a new mechanism for the electromagnetic
radiation by a neutrino moving in the background matter. We have termed this
radiation the \ ``spin light of neutrino" \ ($SL\nu$) in matter
\cite{LobStudPLB03}. The term ``spin light" was used \cite{SL_theor} for
designation of the magnetic-dependent term in the radiation of an electron in a
magnetic field. The $SL\nu$ effect also studied in the cases when
electromagnetic and gravitational fields also present in matter
\cite{DvoGriStudIJMPD05}. Here we should like to mention that the considered
$SL\nu$ is indeed a new type of electromagnetic radiation of a neutrino that
can be emitted by the particle in matter. This radiation mechanism has never
been considered before. As it was mentioned in our first papers on this subject
\cite{LobStudPLB03}, the $SL\nu$ in matter can not be considered as the
neutrino Cherenkov radiation in matter because it can exist even when the
emitted photon refractive index is equal to unit. The $SL\nu$ radiation is due
to radiation of the neutrino by its own rather then radiation of the background
particles.

As it was clear from the very beginning \cite{LobStudPLB03}, the $SL\nu$ is a
quantum phenomenon by its nature and later on we elaborated
\cite{StuTerPLB05GriStuTerPLB05GriStuTerG&C05} the quantum theory of this
radiation (see also \cite{LobPLB05LobDAN05}). To put it on a solid ground, we
of have elaborated a rather powerful method that implies the use of the exact
solutions of the modified Dirac equation for the neutrino wave function in
matter.

Recently we have spread this developed method of the ``exact solutions" to
description of an electron moving matter
\cite{StuJPA06,StuAFB06,GriShiStuTerTro12LomConGriStuTerTroShiFizJ07} and
derived the modified Dirac equation for an electron moving in matter and found
its solutions. On the basis of this exact solution of this equation we have
considered a new mechanism for the electromagnetic radiation that can be
emitted by an electron in the background matter. This mechanism is similar to
the $SL\nu$ in matter and we termed it the ``spin light of electron" in matter
\cite{StuJPA06}.

As it was shown in \cite{StuTerPLB05GriStuTerPLB05GriStuTerG&C05, StuJPA06,
StuAFB06, GriShiStuTerTro12LomConGriStuTerTroShiFizJ07}, in the case of the
standard model interactions of electron neutrinos and electrons with matter
composed of neutrons, the corresponding modified Dirac equations for each of
the particles can be written in the following form:
\begin{equation}\label{new_e}
\Big\{ i\gamma_{\mu}\partial^{\mu}-\frac{1}{2}
\gamma_{\mu}(c_l+\gamma_{5}){\widetilde f}^{\mu}-m_l \Big\}\Psi^{(l)}(x)=0,
\end{equation}
where for the case of neutrino $m_l=m_\nu$ and $c_l=c_{\nu}=1$, whereas for
electron $m_l=m_e$ and $c_l=c_e=1-4\sin^{2}\theta_{W}$. For unpolarized matter
$\widetilde{f}^{\mu}=\frac{G_{F}}{\sqrt{2}}(n_n,n_n{\bf v}),$ $n_n$ and
$\mathbf v$ are, respectively, the neutron number density and overage speed.
The solutions of these equations are as follows,
\begin{equation}
\Psi^{(l)}_{\varepsilon, {\bf p},s}({\bf r},t)=\frac{e^{-i(
E^{(l)}_{\varepsilon}t-{\bf p}{\bf r})}}{2L^{\frac{3}{2}}}
\left(%
\begin{array}{c}{\sqrt{1+ \frac{m_l}{ E^{(l)}_
{\varepsilon}-c\alpha_n m_l}}} \ \sqrt{1+s\frac{p_{3}}{p}}
\\
{s \sqrt{1+ \frac{m_l}{ E^{(l)}_{\varepsilon}-c\alpha_n m_l}}} \
\sqrt{1-s\frac{p_{3}}{p}}\ \ e^{i\delta}
\\
{  s\varepsilon \eta\sqrt{1- \frac{m_l}{ E^{(l)}_{\varepsilon}-c\alpha_n m_l}}}
\ \sqrt{1+s\frac{p_{3}}{p}}
\\
{\varepsilon \eta\sqrt{1- \frac{m_l}{ E^{(l)}_{\varepsilon}-c\alpha_n m_l}}} \
\ \sqrt{1-s\frac{p_{3}}{p}}\ e^{i\delta}
\end{array}
\right).
\end{equation}
where the energy spectra are
\begin{equation}\label{Energy_e}
  E_{\varepsilon}^{(l)}=
  \varepsilon \eta \sqrt{{{\bf p}}^{2}\Big(1-s\alpha_n
  \frac{m_l}{p}\Big)^{2}
  +{m}^2} +c_l {\alpha}_n m_l, \ \ \alpha_n=\pm\frac{1}{2\sqrt{2}}
  {G_F}\frac{n_n}{m_l}.
\end{equation}
Here $p$, $s$ and $\varepsilon$ are the particles momenta, helicities and signs
of energy, ``$\pm$" corresponds to $e$ and $\nu_e$. The value
$\eta=$sign$\big(1-s\alpha_n\frac{m_l}{p}\big)$ is introduced to provide a
proper behavior of the neutrino wave function in the hypothetical massless
case.

It should be pointed out  that the derived modified Dirac equations for a
neutrino and electron in matter and their  exact solutions obtained  establish
an effective method for investigation of different phenomena that can arise
when the particles move in dense media (for more details see \cite{StuAFB06}),
including the cases peculiar for astrophysical and cosmological environments.
For example, effects of the Dirac neutrino reflection and trapping, as well as
neutrino-antineutrino annihilation and neutrino pair creation in matter at the
interface between two media with different densities can be considered on this
basis (see \cite{GriStuTerPAN06} and references therein).

Using the exact solutions of the above mentioned Dirac equations for a neutrino
and electron we have performed detailed investigations of the $SL\nu$ and $SLe$
in matter. In particular,  in the case of ultra-relativistic neutrinos ($p\gg
m$) and a wide range of the matter density parameter $\alpha$  for the total
rate of the $SL\nu$ we obtained \cite{StuTerPLB05GriStuTerPLB05GriStuTerG&C05}
\begin{equation}\label{gamma_nu}
\Gamma_{SL\nu} = 4 \mu_\nu ^2 \alpha ^2 m_\nu^2 p,  \ \ \ \ \ \ \ \
  {m_\nu}/{p} \ll \alpha \ll {p}/{m_\nu}.
\end{equation}

The main properties of the $SL\nu$ investigated in
\cite{LobStudPLB03,DvoGriStudIJMPD05, StuTerPLB05GriStuTerPLB05GriStuTerG&C05}
can be summarized as follows:
 1) a neutrino with nonzero mass and magnetic moment when
moving in dense matter can emit spin light; 2) in general,  $SL\nu$ in matter
is due to the dependence of the neutrino dispersion relation in matter on the
neutrino helicity; 3) the $SL\nu$ radiation rate and  power depend on the
neutrino magnetic moment and energy, and also on the matter density; 4)  the
matter density parameter $\alpha$, that depends on the type of neutrino and
matter composition,  can be negative; therefore the types of initial and final
neutrino (and antineutrino) states, conversion between which can effectively
produce the $SL\nu$ radiation, are determined by the matter composition; 5) the
$SL\nu$ in matter leads to the neutrino-spin polarization effect; depending on
the type of the initial neutrino (or antineutrino) and matter composition the
negative-helicity relativistic neutrino (the left-handed neutrino $\nu_{L}$) is
converted to the positive-helicity neutrino (the right-handed neutrino
$\nu_{R}$) or vice versa; 6) the obtained expressions for the $SL\nu$ radiation
rate and power exhibit non-trivial dependence on the density of matter and on
the initial neutrino energy; the $SL\nu$ radiation rate and power are
proportional to the neutrino magnetic moment squared which is, in general, a
small value and also on the neutrino energy, that is why the radiation
discussed can be effectively produced only in the case of ultra-relativistic
neutrinos; 7) for a wide range of matter densities the radiation is beamed
along the neutrino momentum, however the actual shape of the radiation spatial
distribution may vary from projector-like to cap-like, depending on the
neutrino momentum-to-mass ratio and the matter density; 8) in a wide range of
matter densities the $SL\nu$ radiation is characterized by total circular
polarization; 9) the emitted photon energy is also essentially dependent on the
neutrino energy and matter density; in particular, in the most interesting for
possible astrophysical and cosmology applications case of ultra-high energy
neutrinos, the average energy of the $SL\nu$ photons is one third of the
neutrino momentum. Considering  the listed above properties of the $SL\nu$ in
matter, we argue that this radiation can be produced by high-energy neutrinos
propagating in different astrophysical and cosmological environments.

Performing the detailed study of the $SLe$ in neutron matter
\cite{GriShiStuTerTro12LomConGriStuTerTroShiFizJ07} we have found for the total
rate
\begin{equation}\label{gamma_e}
\Gamma_{SLe}=e^2 m_e^2/(2p)\left[\ln\big({4\alpha_n p}/{m_e}\big)-
    {3}/{2}\right],  \ \ \ \ {m_e}/{p}\ll\alpha_n\ll {p}/{m_e},
\end{equation}
where it is supposed that $\ln \frac{4\alpha _n p}{m_e} \gg 1$. It was also
found that for relativistic electrons the emitted photon energy can reach the
range of gamma-rays. Furthermore,  the electron can loose nearly the whole of
its initial energy due to the $SLe$ mechanism.

Several aspects of the background plasma effects in the $SL\nu$ radiation
mechanism have been discussed in
\cite{StuTerPLB05GriStuTerPLB05GriStuTerG&C05}. Recently this problem has been
also considered in \cite{KuzMikhMPL06IJMPA07} and the total rates of the
$SL\nu$ and $SLe$ in plasma where derived. The final result of
\cite{KuzMikhMPL06IJMPA07} for the $SL\nu$ rate, that accounted for the photon
dispersion in plasma, in the case of ultra-high energy neutrino (i.e., when the
time scale of the process can be much less than the age of the Universe)
exactly reproduces our result (\ref{gamma_nu}) obtained in
\cite{StuTerPLB05GriStuTerPLB05GriStuTerG&C05}. At the same time, the $SLe$
total rate given by eq. (65) in the second paper of \cite{KuzMikhMPL06IJMPA07}
in the leading logarithmic term confirms our result  (\ref{gamma_e}) obtained
in  \cite{GriShiStuTerTro12LomConGriStuTerTroShiFizJ07}.

Recently we have applied the developed method of exact solutions of quantum
wave equations in the background matter to a particular case when a neutrino is
propagating in a rotating medium of constant density \cite{GriSavStuIzvVuz07}.
Suppose that the neutrino propagates inside a uniformly rotating medium
composed of neutrons. This can be considered for modelling of neutrino
propagation inside a rotating neutron star. The corresponding modified Dirac
equation for the neutrino wave function is given by (\ref{new_e}) with the
potential $\widetilde{f}^{\mu}$ that accounts for the medium rotation. The
equation can be solved in the considered case and for the energy spectrum of
the relativistic active left-handed neutrinos with vanishing mass we have
obtained
\begin{equation}
\label{energy_L} p_0 = \sqrt{p_3^2 + 2 \gamma N} - {G}_F n/\sqrt{2}, \ \
\gamma=G_F \omega n/\sqrt{2}, \ \ N=0,1,2,... ,
\end{equation}
where $\omega$ is the angular frequency of the star rotation. The energy
depends on the neutrino momentum component $p_3$ along the rotation axis of
matter and the quantum number $N$ that determines the value of the neutrino
momentum in the orthogonal plane. Thus, it is shown that the transversal motion
of an active neutrino is quantized  very much like an electron energy is
quantized in a constant magnetic field forming the Landau energy levels. From
these properties of the neutrino energy spectrum we predict that there is an
effect of trapping neutrinos with the correspondent energies inside rotating
dense stars.

 The two of the authors (A.G. and A.S.) are thankful to Anatoly Efremov and
 Oleg Teryaev for the invitation to attend the XII Workshop on High
 Energy Spin Physics and for the kind hospitality provided in Dubna.

\end{document}